\documentclass[useAMS,usenatbib,usegraphicx]{mn2e}

\newcommand{\expect}[1]{\langle #1 \rangle}     
\newcommand{\lexpect}[1]{\left\langle #1 \right\rangle}  
\newcommand{\Mpc}{$h^{-1}$\,Mpc}
\newcommand{\iMpc}{$h$\,Mpc$^{-1}$}

\newcommand{\bk}{{\bmath k}}
\title[Information in the non-linear power spectrum]{Information
  content of the non-linear matter power spectrum}
\author[C. D. Rimes and A. J. S. Hamilton]
	   {Christopher D. Rimes$^1$\thanks{E-mail: rimes@colorado.edu, %
		 Andrew.Hamilton@colorado.edu} %
		 and Andrew J. S. Hamilton$^{1, 2}$$^\star$\\
		 $^1$JILA, University of Colorado, 440 UCB, Boulder, CO 80309-0440, %
		 U.S.A.\\
		 $^2$Department of Astrophysical and Planetary Sciences, %
		 University of Colorado, 391 UCB, Boulder, CO 80309-0391, U.S.A.}

\begin{document}

\date{Accepted 2005 April 11. Received 2005 April 1; in original form 2005 February 3}

\pagerange{\pageref{firstpage}--\pageref{lastpage}} \pubyear{2004}

\maketitle

\label{firstpage}

\begin{abstract}
We use an ensemble of N-body simulations of the currently favoured
(concordance) cosmological model to measure the amount of information contained
in the non-linear matter power spectrum
about the amplitude of the initial power spectrum.  Two surprising results
emerge from this study: (i) that there is very little independent information
in the power spectrum in the translinear regime ($k \simeq 0.2$--0.8~\iMpc\ at
the present day) over and above the information at linear scales and (ii) that
the cumulative information begins to rise sharply again with increasing
wavenumber in the non-linear regime.  In the fully non-linear regime, the
simulations are consistent with no loss of information during translinear and
non-linear evolution.  If this is indeed the case then the results suggest a
picture in which translinear collapse is very rapid, and is followed by a
bounce prior to virialization, impelling a wholesale revision of the HKLM-PD
formalism.
\end{abstract}

\begin{keywords}
cosmology: theory -- large-scale structure of Universe.
\end{keywords}

\section{Introduction}

Measurements of galaxy clustering play a key role in the quest to accurately
determine cosmological parameters.  As pointed out by
\citet*[1999\nocite{EHT99}]{EHT98}, constraints from galaxy clustering directly
complement those provided by the cosmic microwave background (CMB), enabling
the breaking of parameter degeneracies which affect both types of observation
when considered in isolation.  The power of combining datasets in this way has
recently been demonstrated by several groups (\citealt{Seljak04};
\citealt{Tegmark04a}; \citealt{Efstathiou02}) and the existence of a so-called
`concordance' cosmological model is largely thanks to this type of effort.

On large scales, the power spectrum of galaxy clustering traces the spectrum of
primordial density fluctuations on those scales and is therefore a direct test
of early-universe models.  Many such models (including the simplest forms of
the inflationary scenario) predict perturbations to the density field that are
Gaussian random distributed, and to date this generic prediction has remained
consistent with observation \citep{Komatsu03}.  For Gaussian fluctuations the
power spectrum contains all possible statistical information about the
perturbations.

At smaller scales, where much of the information in galaxy surveys lies, the
extent to which the linear power spectrum can be recovered from the non-linear
power spectrum remains unknown.  The success of the HKLM-PD formalism
(\citealt{HKLM91}; \citealt[1996\nocite{PD96}]{PD94}) in relating the
non-linear power spectrum to the linear one suggests that non-linear evolution
preserves at least some of the information in the power spectrum, albeit
transported from larger to smaller comoving scales by the process of
gravitational collapse.  On the other hand, the simulations of \citet*{MWP99}
indicate that non-linear evolution washes out baryonic wiggles in the power
spectrum, suggesting that some information is perhaps irreversibly destroyed.

The purpose of the present paper is to investigate quantitatively, using N-body
simulations, whether or not non-linear evolution preserves information in the
matter power spectrum.


\section{Information}
\label{information}

\begin{figure}
\centerline{\includegraphics[width=8cm]{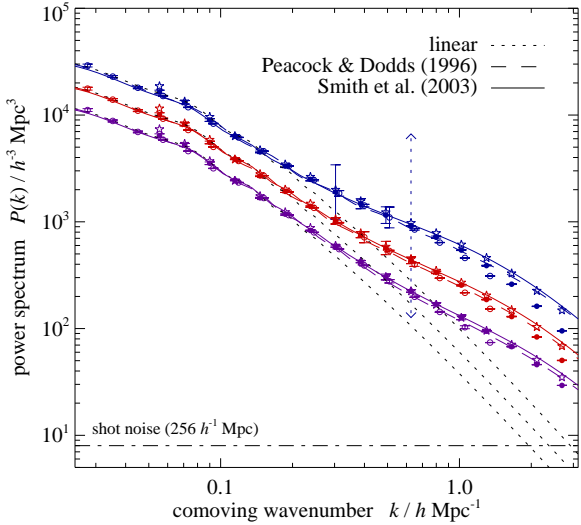}}
\caption{Evolution of the non-linear power spectrum.  Points with error bars
  show the mean power spectrum at three epochs (bottom to top: $a=0.5$, 0.67
  and 1) in the PM simulations.  Open points are averages over the 256~\Mpc\
  simulations and filled points are the same for the 128~\Mpc\ simulations.
  The error bars are statistically independent.  The dashed error bar extends
  beyond the bounds of the axes plotted.  Stars are the results from the 25
  higher resolution ART simulations (shown without error bars for clarity).
  The linear power spectrum is shown by the dotted curves in each panel.  The
  solid and dashed curves are, respectively, from the fitting formulae of
  \citet{Smith03} and \citet{PD96}. The dot-dashed line marks the level of the
  shot noise in the 256~\Mpc\ boxes.
  \label{fig: power}}
\end{figure}

We measure the Fisher information \citep*{TTH97} $I$ in the log of the
amplitude $A$ of the initial (post-recombination) matter power spectrum:
\begin{equation}
\label{eq: I}
  I \equiv - \lexpect{ \frac{\partial^2\ln\mathcal{L}}{\partial\ln{A}^2} }.
\end{equation}
Here $\mathcal{L}$ denotes the likelihood, which for Gaussian fluctations takes
the form $\mathcal{L} \propto | C |^{-1/2} \exp\left(- \frac{1}{2} \bdelta
C^{-1} \bdelta \right)$, where $\bdelta$ is the observed data vector of
fluctuations, and $C$ is their expected covariance matrix.  If fluctuations are
statistically homogeneous and isotropic, then each Fourier mode $\delta_\bk$ is
independently Gaussianly distributed.  The variance of Fourier modes, the
diagonal elements of the diagonal covariance matrix, constitute the power
spectrum $\expect{|\delta_\bk|^2} \propto P(k)$.

Thus, for Gaussian fluctuations, the likelihood depends on parameters only
through the power spectrum $P(k)$, and the information $I$ defined by
equation~(\ref{eq: I}) can be written
\begin{equation}
\label{eq: IPk}
  I = - \left\langle
      \sum_{k,k^\prime}
      \frac{\partial\ln P(k)}{\partial\ln A}
      \frac{\partial^2\ln\mathcal{L}}{\partial\ln P(k) \, \partial\ln P(k^\prime)}
      \frac{\partial\ln P(k^\prime)}{\partial\ln A}
      \right\rangle.
\end{equation}
For simplicity, in this paper we use results from simulations only at scales
where the shot noise contribution to the power is subdominant, so that $P(k)$
in equation~(\ref{eq: IPk}) can be regarded as the cosmic variance.

During linear evolution, the partial derivatives of the log power with respect
to log amplitude in equation~(\ref{eq: IPk}) are just unity, $\partial\ln P(k)
/ \partial\ln A = 1$.  A short calculation from the Gaussian likelihood
function shows that, for Gaussian fluctuations, the information $I$ of
equation~(\ref{eq: IPk}) equals half the number $N$ of Gaussian modes:
\begin{equation}
  I = N/2.
\end{equation}
Here $\delta_\bk$ and its complex conjugate $\delta_{-\bk}$ are counted
as contributing two distinct modes, the real and imaginary parts of
$\delta_\bk$.

At non-linear scales, we continue to define the information $I$ in the
non-linear power spectrum $P(k)$ by equation~(\ref{eq: IPk}).  Clearly, there
is a mapping from the initial linear power spectrum $P_{\rm L}(k)$ to the
non-linear power spectrum $P(k)$: to find it, just do an $N$-body simulation
(the cosmic variance in the non-linear power spectrum should be negligible if
the simulation is large enough).  On the other hand, it is not clear a priori
whether an inverse mapping exists.  If it does, then the Fisher information
$I$, equation~(\ref{eq: IPk}), in the non-linear power spectrum should equal
that in the initial linear power spectrum: information is preserved.
Conversely, if an inverse mapping does not exist, then the Fisher information
in the non-linear power spectrum should be less than that in the initial power
spectrum: non-linear evolution destroys information.

The definition~(\ref{eq: IPk}) of information involves partial derivatives
$\partial\ln P(k) / \partial\ln A$ of the log of the non-linear power with
respect to the log of the initial, linear amplitude.  In an $N$-body
simulation, increasing the initial amplitude is equivalent to evolving the
simulation further in time.  Thus the desired partial derivatives can be
measured simply by comparing the amplitudes of non-linear power $P(k)$ at
successive epochs.  It is this property that makes the information in the log
amplitude especially convenient to measure from simulations: there is no need
to perform simulations with different values of cosmological parameters.

The other factor in the definition~(\ref{eq: IPk}) of information is the second
derivative of the log likelihood with respect to the log non-linear powers.
This factor is the Hessian of the vector $\ln\!P(k)$ of log non-linear powers,
the expectation value of which is the Fisher matrix with respect to the log
powers.  Since each $P(k)$ involves an expectation over many modes
$\delta_\bk$, it is reasonable to invoke the central limit theorem to assert
that estimates of power will be distributed in a Gaussian fashion about the
expectation value, in which case this factor is approximately equal to the
inverse of the covariance matrix of power spectrum estimates.  This assertion
holds even if the density field itself is non-Gaussian.  The reliability of the
approximation can be checked at linear scales (where there are fewest modes so
the central limit theorem is least likely to apply), where the Fisher matrix
should be diagonal, with diagonal elements equal to half the number of modes in
each wavenumber bin.

\begin{figure*}
\includegraphics[width=16cm]{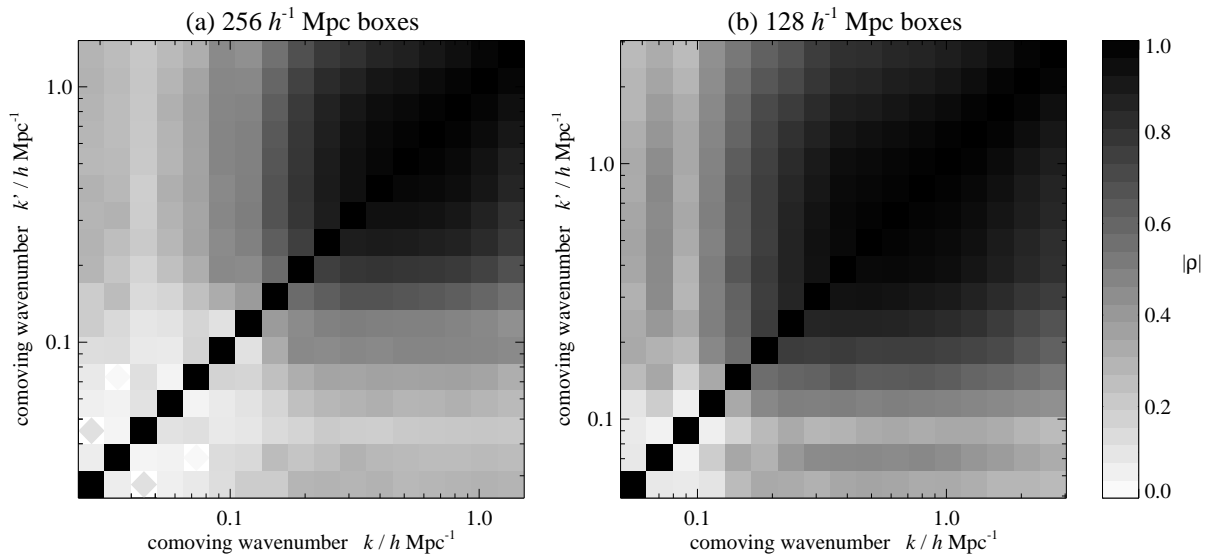}
\caption{Correlation matrices of estimates of the non-linear power spectrum for
  (a) the 256~\Mpc\ boxes and (b) the 128~\Mpc\ boxes.  Positive correlations
  are shown by completely filled bins while diamonds denote anti-correlations.
  Greyscale represents the magnitude of the correlations, ranging from 0 (no
  correlation) to 1 (perfect correlation).
  \label{fig: correlations}}
\end{figure*}



\section{Simulations}
\label{simulations}

We generated 400 random realizations of a cubic region of the universe
256~\Mpc\ on the side.  A further 200 realizations with a box size of 128~\Mpc\
were used to isolate numerical effects close to the mesh scale.  The
cosmological model used was the `vanilla-lite' model of \citet[second-last
column of their table 4]{Tegmark04a}: $\Omega_{\rm M}=0.29$,
$\Omega_\Lambda=0.71$, baryon fraction $f_{\rm b}\equiv\Omega_{\rm
b}/\Omega_{\rm M}=0.16$, $h =0.71$ and $\sigma_8=0.97$.  The matter transfer
function was calculated using the fitting formula of \citet{EH98}.  The boxes
were evolved using a particle-mesh (PM) code with $128^3$ particles and a
$256^3$ force mesh.  Adaptive time-stepping was used to control the force
errors; a typical run required between 800 and 1400 steps, with the 128\Mpc\
boxes requiring more, on average, because of the higher degree of clustering.
We also ran 25 realizations of a 128~\Mpc\ box at higher resolution using the
adaptive mesh refinement code ART \citep*{KKK97} with a 128$^3$ root mesh and 3
levels of refinement.  The initial conditions for these simulations were set up
using the GRAFIC package with the same cosmological parameters as above.
Although the small number of realizations is not sufficient to yield a precise
estimate of the covariance matrix -- we find that at least several hundred
simulations are required to achieve convergence -- they serve to confirm the
results of the much larger set of PM simulations on small scales.

The evolution of the non-linear power spectrum is illustrated by Fig.~\ref{fig:
power}, which shows the mean, shot noise-subtracted, power spectrum of the
simulations at three epochs.  The power spectrum was measured by calculating
the density field on a 256$^3$ cubic mesh, using a cloud-in-cell scheme, and
binning the resulting Fourier amplitudes in radial bins in $k$-space.  For the
ART simulations, `chaining' \citep{Jenkins98} was used to reach small scales.
Smoothing due to the mass assignment scheme was corrected for by dividing by
the square of the Fourier transform of the window function, prior to
subtracting the shot noise contribution \citep{Smith03}.  In the following
analysis, we use only wavenumbers a factor of at least two away from the
Nyquist frequency of the Fourier transform grid to avoid problems resulting
from the uncertain mass assignment correction and aliasing of power from
smaller scales.  The power spectra from the two different box sizes agree well
up to approximately 3 times the scale of the force mesh of the larger boxes,
but at smaller scales the power in the PM simulations is systematically
underestimated relative to the higher resolution ART simulations.  The latter
agree well with fits to the results of \citet{Smith03} and \citet{PD96}.  The
error bars in Fig.~\ref{fig: power}, which are statistically independent, are
actually attached to the uncorrelated band powers described below.


In Fig.~\ref{fig: correlations} we plot the correlation coefficients between
estimates of log-power in each pair of wavenumber bins, measured from the
simulations.  This is simply the covariance matrix scaled so that the diagonal
elements are identically unity and has the advantage that, whereas the elements
of the covariance matrix are smaller for larger simulation volumes, the
correlation co-efficients are independent of box size.  The results for the 200
128~\Mpc\ PM simulations and the 25 ART simulations are consistent, so we
combine them into a single covariance matrix for this box size.  Our results
confirm those of \citet{MW99}, who found that (i) correlations between
neighbouring band powers grow rapidly in the translinear regime, approaching
100\% at non-linear wavenumbers and (ii) even for pairs of bands in which only
one of the pair is non-linear there are significant correlations, of the order
of 20--40\%.  The results from the two different box sizes are again broadly
consistent.

\begin{figure*}
\centerline{\includegraphics[width=15cm]{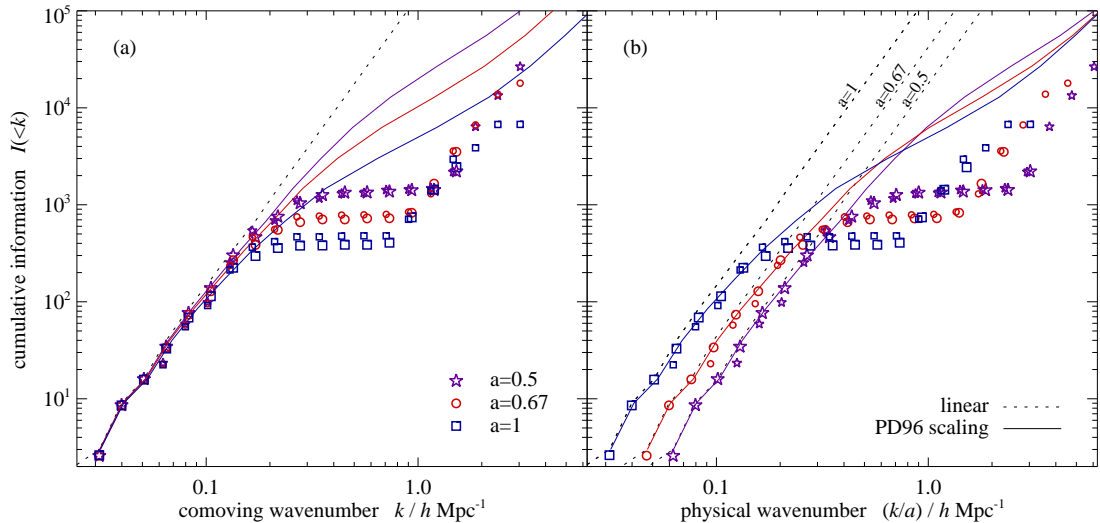}}
\caption{Cumulative information in the non-linear power spectrum at three
  epochs ($a=0.5$, 0.67 and 1) as a function of (a) comoving and (b) physical
  wavenumber.  Large symbols are points derived from the 256~\Mpc\ boxes and
  small symbols are from the 128~\Mpc\ boxes (PM+ART).  The results for the
  128\Mpc\ boxes have been shifted vertically by a factor 8 to account for the
  higher density of modes at a given comoving $k$.  The dotted lines show the
  information in the linear power spectrum.  The solid curves are the result of
  applying the \citet{PD96} wavenumber scaling to the linear information.
  \label{fig: info}}
\end{figure*}

We assign information to each wavenumber bin by defining a set of uncorrelated
band-power estimates:
\begin{equation}
  \frac{\hat{B}(k)}{P(k)} = 
          \sum_{k^\prime} W(k,k^\prime)
                  \frac{\hat{P}(k^\prime)}{P(k^\prime)}
\label{eq: band powers}
\end{equation}
(we use hats to denote measurements from individual simulations and hatless
symbols for averages over all simulations).  $W(k,k^\prime)$ is a decorrelation
matrix \citep{HT2000}.  Of the infinity of possible schemes for decorrelating
the power spectrum, we use here the upper Cholesky decomposition of the Fisher
matrix of the scaled power spectrum, suitably normalized, as our decorrelation
matrix.  We experimented with several decorrelation schemes and this was the
only one that reproduced the expected amount of information in the linear
regime, where it is reasonable to expect that information is conserved.
Mathematically, it is equivalent to taking a matrix composed of all the
elements of the covariance matrix up to some wavenumber $k_{\rm max}$,
inverting this matrix and summing all the elements of the resulting Fisher
matrix to arrive at a measure of the accumulated information, $I(<k_{\rm
max})$, up to that wavenumber.  Scaling both sides of equation~(\ref{eq: band
powers}) by $P(k)$ guarantees that the mean of the uncorrelated band power
estimates at each wavenumber is equal to the mean power spectrum.  It is the
errors on the decorrelated band powers that are plotted in Fig.~\ref{fig:
power}.  Notice that some of the band powers have error bars much larger than
the actual scatter of the data points; this simply implies that those points
contain almost no independent information.


\section{Discussion}
\label{}

Fig.~\ref{fig: info} presents the essential results of this paper.  It shows
the cumulative information as a function of both comoving and physical
wavenumber at three epochs.  This plot reveals two striking and unexpected
features.  The first is how flat the cumulative information becomes in the
translinear regime ($k \simeq 0.2$--0.8~\iMpc\ at $a=1$).  The flatness appears
consistently at all epochs, in both box sizes and using two different
simulation codes.  The results from the 25 ART simulations are consistent,
within the scatter, with the larger suite of 128~\Mpc\ PM simulations so we
have combined the results in a single curve for this box size, using the power
spectrum from the ART simulations, which are more accurate on small scales, to
calculate the partial derivatives $\partial\ln P(k) / \partial\ln A$ in
equation (\ref{eq: I}).  There are two possible reasons for the flatness of
cumulative information: either information is being lost from the power
spectrum rather abruptly as structures enter the translinear regime and begin
to collapse, or else information is flowing rapidly from large to small scales.
The HKLM-PD formalism predicts that information should indeed flow from large
to small scales as structures collapse but, as Fig.~\ref{fig: info} shows, if
information is preserved then the flow of information is far more rapid than
predicted by the PD formula.

The second remarkable feature of the curves in Fig.~\ref{fig: info} is that in
the highly non-linear regime the cumulative information begins to rise sharply
again (at $k \simeq 0.8$\iMpc\ for $a=1$).  This upturn occurs consistently at
all epochs, in both box sizes and with both codes, making it unlikely to be a
numerical effect in the simulations (which ought, at the very least, to scale
with the box size).  The upturn is difficult to explain if information in the
power spectrum is completely destroyed during translinear collapse.  The
possibility remains that during translinear evolution information is
temporarily diverted out of the power spectrum into higher order moments and
that it is somehow restored into the power spectrum at non-linear scales, but
this explanation seems contrived and we do not explore it further.

The definitive test for whether information is being destroyed is to look at
the cumulative information in the highly non-linear regime.  At highly
non-linear scales, structures virialize and therefore cease to collapse
rapidly.  The HKLM-PD formalism assumes the stable clustering hypothesis: that
following virialization structures remain of fixed size in physical
co-ordinates.  Probably, the assumption of stable clustering is not precisely
true (e.g.\ \citealt{Padmanabhan96}; \citealt{Smith03}).  Nevertheless, the
collapse or expansion of structures in the virialized regime is much slower
than the rapid dynamic collapse that takes place in the translinear regime.

It follows that, if translinear and non-linear evolution preserve information
in the power spectrum, then the cumulative information at a given fixed
physical scale in the highly non-linear regime should remain constant.  In
other words, the cumulative information at different epochs, plotted as a
function of \emph{physical} wavenumber $k / a$, should asymptote to a single
common line in the highly non-linear regime, as it does in the PD formula.
Conversely, if evolution destroys information, then the cumulative information
at a fixed non-linear physical scale should decrease systematically with time.
Fig.~\ref{fig: info}(b) shows that shows that the cumulative information in the
non-linear power spectrum
at the smallest physical scales that can be reliably measured, is consistent,
within the uncertainties, with no loss of information.  What is more surprising
is that on small scales, the cumulative information at the two later epochs
\emph{exceeds} that at the same physical scale at $a=0.5$, although there are
indications that all three curves will eventually asymptote to the same value
in the highly non-linear regime.  This temporary increase of information
suggests that structures bounce back prior to virialization.  At $a=1$, for
example, structures at $k \simeq 0.2$--0.8~\iMpc\ are in rapid collapse,
following turnaround.  At smaller scales, structures have bounced and are
actually expanding, before becoming fully virialized at $k \simeq 3$~\iMpc.
 
 
The simulations presented here are consistent with the hypothesis that
non-linear evolution largely preserves information in the power spectrum.  If
this is true, then a wholesale revision of the HKLM-PD formalism is needed.
The simulations suggest rapid translinear collapse followed by a bounce and
subsequent virialization, in contrast to the rather gentle behaviour predicted
by the HKLM-PD formalism.

This rapid collapse can be construed as supporting the alternative picture of
non-linear evolution put forward by the halo model (e.g.\ \citealt{MF2000};
\citealt{PS2000}; \citealt{Seljak2000}).  In the halo model, the cosmic density
field is treated as a set of discrete, collapsed dark matter haloes, whose
centres are clustered according to linear theory.  The non-linear contribution
to the matter power spectrum is determined by the density profiles of the
individual haloes.  Implicit in this picture is the assumption that the
transition between the linear and virialized regimes is an abrupt one.  Our
direct measurements of the flow of information from large to small scales
confirm that this indeed appears to be the case.

The results reported in this paper have several other practical implications
beyond those relevant to analytic models of non-linear evolution.  Firstly,
if non-linear evolution completely preserves information in the power spectrum,
then information about baryonic wiggles is preserved.  Different and better
simulations than carried out for this paper will be necessary to test what
actually happens to baryonic wiggles.  Secondly, if the translinear collapse to
small scales is indeed as rapid as suggested by the simulations in this paper,
then baryonic wiggles should be stretched over translinear scales much more
than had previously been anticipated.  Thirdly, if the translinear power
spectrum contains little additional information over and above that in the
linear power spectrum, then efforts to measure power at translinear scales may
not be as rewarding, in the sense of refining estimates of cosmological
parameters, as might have been anticipated.  It should be emphasized that one
should not interpret our results as implying that the translinear regime
contains no information; rather, the information in the translinear regime is
degenerate with that in the linear regime.

In future work it will be interesting not only to test the results of the
present paper at smaller scales, but also to investigate the extent to which
non-linear evolution does or does not preserve information about other
cosmological quantities, such as the primordial spectral index, or the baryon
fraction.


\section*{Acknowledgements}

We are grateful to Nick Gnedin for providing the particle-mesh code used in
this work and for many useful discussions.  We also thank Anatoly Klypin and
Andrey Kravtsov for making the MPI implementation of ART available to us and
for help with its application.  The referee's comments prompted us perform
further investigations which added significantly to our understanding of the
results presented here.  This work was supported by NSF grant AST-0205981 and
by NASA ATP award NAG5-10763.  GRAFIC is part of the COSMICS package
(http://arcturus.mit.edu/cosmics), which was developed by Edmund Bertschinger
under NSF grant AST-9318185.  Some of the simulations used in this work were
performed at the San Diego Supercomputer Center using resources provided by the
National Partnership for Advanced Computational Infrastructure under NSF
cooperative agreement ACI-9619020.

\bibliographystyle{my_mn2e}
\bibliography{info}

\appendix

\bsp

\label{lastpage}

\end{document}